# Moiré Excitons in Van der Waals Heterostructures

Kha Tran[1], Galan Moody[2], Fengcheng Wu[3], Xiaobo Lu[4], Junho Choi[1], Akshay Singh[1,*], Jacob Embley[1], André Zepeda[1], Marshall Campbell[1], Kyounghwan Kim[5], Amritesh Rai[5], Travis Autry[2], Daniel A. Sanchez[6], Takashi Taniguchi[7], Kenji Watanabe[7], Nanshu Lu[6,8], Sanjay K. Banerjee[5], Emanuel Tutuc[5], Li Yang[4], Allan H MacDonald[1], Kevin L. Silverman[2], and Xiaoqin Li[1,8+]

[1]Department of Physics and Center for Complex Quantum Systems, The University of Texas at Austin, Austin, TX 78712, USA.
[2]National Institute of Standards & Technology, Boulder, CO 80305, USA.
[3]Materials Science Division, Argonne National Laboratory, Argonne, IL 60439, USA.
[4]Department of Physics, Washington University in St. Louis, St. Louis, Missouri 63136, USA.
[5]Microelectronics Research Center, Department of Electrical and Computer Engineering, The University of Texas at Austin, Austin, TX 78758, USA.
[6]Department of Aerospace Engineering and Engineering Mechanics, The University of Texas at Austin, Austin, TX 78712, USA.
[7]National Institute of Material Science, 1-1 Namiki, Tsukuba, Ibaraki 205-0044, Japan.
[8]Texas Materials Institute, The University of Texas at Austin, Austin, TX 78712, USA.
[*]Present address - Department of Material Science and Engineering, Massachusetts Institute of Technology, Cambridge, MA 02139, USA.

## Abstract

In van der Waals (vdW) heterostructures formed by stacking two monolayer semiconductors, lattice mismatch or rotational misalignment introduces an in-plane moiré superlattice. While it is widely recognized that a moiré superlattice can modulate the electronic band structure and lead to novel transport properties including unconventional superconductivity and insulating behavior driven by correlations, its influence on optical properties has not been investigated experimentally. We present spectroscopic evidence that interlayer excitons are confined by the moiré potential in a high-quality $MoSe_2/WSe_2$ heterobilayer with small rotational twist. A series of interlayer exciton resonances with either positive or negative circularly polarized emission is observed in photoluminescence, consistent with multiple exciton states confined within the moiré potential. The recombination dynamics and temperature dependence of these interlayer exciton resonances are consistent with this interpretation. These results demonstrate the feasibility of engineering artificial excitonic crystals using vdW heterostructures for nanophotonics and quantum information applications.

## Main Text

Recent advances in isolating and stacking monolayers of van der Waals (vdW) materials have provided a new approach for creating quantum systems in the ultimate two-dimensional limit[1,2]. In vdW materials, the usual constraint of lattice matching between adjacent layers is lifted. Furthermore, the twist angle between two vdW-bonded layers can be adjusted arbitrarily, in contrast to conventional epitaxially grown heterostructures in which the orientation of adjacent layers is fixed by the crystal axes. These unique properties of vdW heterostructures present new possibilities for engineering electronic band structure and optical properties via an in-plane moiré superlattice. In twisted graphene bilayers, for example, a moiré superlattice has led to the observation of unconventional superconductivity[3] and Hofstadter-butterfly spectra in the presence of a strong magnetic field[4-6].

In vdW semiconductors, such as transition metal dichalcogenides (TMDs), stacking two different monolayers forms a heterobilayer (hBL) with strong out-of-plane quantum confinement. Moiré superlattices form either due to lattice constant mismatch or a twist angle between layers, but they are not necessarily present in TMD hBLs. For example, the lattice constants of TMDs that share common chalcogen atoms (*i.e.*, $MoX_2$ and $WX_2$) only differ by ~0.1%. In rotationally aligned $MoSe_2/WSe_2$ hBLs grown by chemical vapor deposition (CVD) methods, the minor lattice distortion in each layer leads to a commensurate atomic alignment and the moiré pattern is absent[7]. In mechanically stacked hBLs, however, a twist angle between the two layers is unavoidable. Thus, a moiré pattern is expected and indeed has been directly imaged with high-resolution transmission electron microscopy[8].

In TMD hBLs with a typical type-II band alignment[9-11], rapid charge transfer[12,13] between the layers following optical excitation leads to emission from the lower-energy interlayer exciton transition[14-21]. Theoretically, multiple interlayer exciton resonances are expected to form due to the lateral confinement from the moiré potential (Fig. 1)[22-24]. The depth of the moiré potential depends on the interlayer coupling. It is predicted to be ~100-200 meV by first-principles calculations, which have been confirmed by scanning tunneling spectroscopy experiments on a rotationally aligned $MoS_2/WSe_2$ bilayer grown by CVD[25]. Such a deep potential is expected to localize interlayer excitons as long as the moiré supercell has a period of 10 nm or larger[22,24]. Direct spatial imaging of excitons confined in the moiré potential is extremely challenging because of the small dimensions of the moiré supercell. If and how the moiré potential manifests in far-field diffraction-limited optical measurements remains an outstanding question. Addressing this question will bring completely new perspectives on engineering the optical properties of vdW heterostructures.

Here, we present spectroscopic evidence of interlayer exciton (*IX*) resonances confined by the moiré potential in a high-quality hexagonal boron nitride (hBN) encapsulated MoSe$_2$/WSe$_2$ hBL. Because the layers are aligned with a small twist angle and the inhomogeneous spectral linewidths are reduced with the capping layers, several nearly equally spaced *IX* resonances are spectrally resolved at low temperature. Upon excitation with σ+ circularly polarized light, the *IX* resonances exhibit alternating co- and cross-circularly polarized emission. We suggest that the alternating polarized emission originates from the atomic-scale spatial variation of the optical selection rules within a moiré supercell. The energy spacing of the resonances and helicity of the emitted light are consistent with calculations of multiple *IX* states confined within a quantum dot-like potential with 100-200 meV lateral confinement, which agrees with the expected moiré potential depth. Time-resolved and temperature-dependent photoluminescence measurements support this assignment of the ground and excited state *IX* excitons. This collection of spectroscopic features can only be explained by taking into account the effects of the moiré potential. Our work will stimulate future studies exploring vdW heterostructures as novel photonic materials and as a platform for engineering an array of single photon emitters, exciton crystals, and quantum phases of excitons.

We first describe conceptually how the moiré potential may give rise to multiple exciton resonances with different optical selection rules, as shown in Fig. 1. In MoSe$_2$/WSe$_2$ hBLs with a small twist angle ≲1°, the exciton Bohr radius is large compared to the monolayer lattice constant but small relative to the moiré supercell (~20-50 nm). Thus, the interlayer exciton can be described as a particle moving in a slowly varying moiré potential that modulates the local bandgap[24,26,27]. Within a moiré supercell, there are three points where the local atomic registration preserves the three-fold rotational symmetry $\hat{C}_3$, as shown in Figs. 1a and 1b. These three sites are denoted by $R_h^h$, $R_h^X$, $R_h^M$ respectively, where $R_h^\mu$ refers to *R*-type stacking with the $\mu$ site of the MoSe$_2$ layer aligning with the hexagon center ($h$) of the WSe$_2$ layer. These high symmetry points are local energy extrema within the moiré supercell where excitons can be localized. In the case of sufficiently deep energy modulation, the moiré pattern can, in principle, provide the lateral confinement to define a two-dimensional array of identical quantum dots (left panel of Fig. 1c).

Another important consequence of the moiré pattern is to impose spatially varying optical selection rules[24,28]. Although the valley degree of freedom is still a good quantum number for interlayer excitons, the optical selection rules of exciton resonances are no longer locked to the valley index as is the case of monolayers. For example, for spin-conserving transitions between tungsten or molybdenum *d*-orbitals at the *K* valley in the first Brillouin zone (see Fig. 1b), the photons generated from recombination must have the same rotational symmetry as the interlayer exciton. As shown in Fig. 1b, an exciton residing directly at site $R_h^h$ ($R_h^X$) only couples to $\sigma_+$ ($\sigma_-$) polarized light. Site $R_h^M$ has a dipole oriented perpendicular to the plane and excitons located at this position do not efficiently couple to light propagating with normal incidence (see supplementary material). These valley optical selection rules are determined not only by atomic quantum numbers, but also by the relative position between tungsten and molybdenum atoms in real space. It is the latter dependence that is responsible for distinct selection rules at different positions. The optical selection rules change continuously in the moiré pattern and are generally elliptically polarized (right panel of Fig. 1c). In the case of a deep moiré potential, quantized exciton states with distinct optical selection rules can contribute to the optical properties, as the experiments and theoretical calculations presented below demonstrate.

To examine the influence of the moiré pattern on interlayer excitons, we perform micro-photoluminescence (PL) measurements on MoSe$_2$/WSe$_2$ hBLs. The samples were prepared following a mechanical exfoliation and transfer process[29,30] described in detail in the methods section. An optical image of an encapsulated hBL with ~1° twist angle is shown in Fig. 2a. Unless otherwise specified, the hBL is held at 15 K and excited with a continuous wave 660 nm laser with a full-width at half-maximum spot size of 1.5 µm. For an uncapped sample, the PL spectrum (dashed curve in Fig. 2b) features intra-layer neutral and charged excitons and a broad *IX* resonance (Fig. 2c) consistent with earlier reports[14,31,32]. When an hBL is encapsulated between hBN layers, four spectrally resolved *IX* resonances emerge (solid curve in Fig. 2b) due to the reduced inhomogeneous linewidth. The *IX* spectral region is replotted in Fig. 3a and fit with four Gaussian functions. The central emission energies extracted from the fits are 1310 meV, 1335 meV, 1355 meV, and 1380 meV. Three representative PL spectra from different locations across the sample are displayed in Fig. 3a. The peak resonance energies are repeatable across different locations on the hBL, with a nearly constant peak spacing of ~20-25 meV (Fig. 3b).

When considering moiré effects on the exciton optical response, multiple *IX* peaks may be indicative of quantized energy levels due to lateral confinement, as predicted in the calculations below. The fact that the states span an energy range of 70 meV suggests that the moiré potential depth is on the order of 100 meV—sufficient to justify the picture of an array of quantum dots. Polarization-resolved PL experiments provide additional compelling evidence in support of this interpretation. Using $\sigma_+$ polarized excitation, we collected co- ($\sigma_+$ detection) and cross-circularly ($\sigma_-$ detection) polarized PL spectra, which are shown in Fig. 3c. We define the circular polarization of emission as $P_c = \frac{I_{co} - I_{cross}}{I_{co} + I_{cross}}$, where $I$ is the measured PL intensity. We plot $P_c$ as a function of energy in Fig. 3d. The *IX* resonances exhibit a variation of $P_c$ between 0.2 and -0.2. A negative $P_c$ indicates that the PL signal with cross-circular polarization is stronger than that from the co-circular polarization. We propose that the alternating co- and cross-circular emission arises from the unique spatial variation of the optical selection rules predicted based on rotational symmetry considerations[24].

To explain the polarization dependence, we consider the population of *IX* transitions in the *K* valley following $\sigma_+$ optical excitation. To relate the observed PL signal to the optical selection rules, we first assume that the above-gap, $\sigma_+$ polarized light optically excites spin-polarized intralayer excitons or free carriers in the MoSe$_2$ and WSe$_2$ monolayers primarily in the *K* valley. This spin-valley locking in TMD monolayers has been established by previous studies[33-37]. Second, we assume that the charge transfer process leading to the *IX* formation conserves the valley and spin index, which is supported by polarization-resolved pump-probe spectroscopy[38]. It then follows that an *IX* state created in the *K* valley following $\sigma_+$ optical excitation emits $\sigma_+$ ($\sigma_-$) polarized light if it is localized near the $R_h^h$ ($R_h^X$) high-symmetry point within the moiré potential landscape (refer to Figs. 1b and 1c). We show in the calculations below for a deep moiré potential that confines excitons at the $R_h^h$ site, the wave functions associated with the quantized exciton states can acquire additional angular momentum and sample the potential landscape in a way that leads to multiple resonances with alternating $\sigma_+$ and $\sigma_-$ light emission—a characteristic consistent with our experimental observations. Because the valley relaxation and charge transfer dynamics can be complex, the above assumptions do not strictly hold; however, the primary effect of deviations from these simple arguments will be creation of *IXs* in the -*K* valleys, which reduces $P_c$ below unity. Nevertheless, we argue that the most plausible explanation for the distinct positive and negative values of $P_c$ associated with different *IX* resonances is spatial variation in optical selection rules within the moiré potential.

In the moiré pattern, the local band gap varies in real space and acts as a periodic potential for excitons. The Bohr radius of *IX* (~1 nm) is significantly smaller than the moiré period, allowing the *IX* to be treated as a composite quasiparticle with a wavepacket moving in the potential. Its center-of-mass (COM) motion is described by

$$H = \hbar\Omega_0 + \frac{\hbar^2 \bm{k}^2}{2M} + \Delta(\bm{r}), \qquad (1)$$

where $\hbar\Omega_0$ is an energy constant, $\hbar^2\bm{k}^2/(2M)$ is the COM kinetic energy, $\Delta(\bm{r})$ is the exciton moiré potential energy, and $M$ is the exciton mass. Because $\Delta(\bm{r})$ is smoothly varying, it can be approximated by the lowest-order harmonic expansion and is then specified completely by one strength and one shape parameter. Since the lowest-energy excitons confined by the potential near the $R_h^h$ site have an *s*-wave symmetric COM wave function and emit $\sigma_+$ light when in the *K* valley, consistent with experiment, we adjust the shape parameter to place the minima of $\Delta(\bm{r})$ at these sites. Near $R_h^h$ sites, $\Delta(\bm{r})$ has the form of a harmonic oscillator with quantized energy levels controlled by the moiré period $a_M$. We take the twist angle to be 1°, resulting in $a_M$ equal to ~19 nm.

Both *K* and -*K* valley excitons are governed by the Hamiltonian in Eqn. (1), but they have different optical responses due to valley-dependent optical selection rules. Below we focus on *K* valley excitons—properties of -*K* valley excitons can be inferred by using time-reversal symmetry. At each spatial position, the interlayer optical matrix element $\bm{J}(\bm{r})$ can be written as $J_+(\bm{r})\hat{\bm{e}}_+ + J_-(\bm{r})\hat{\bm{e}}_-$, where $J_\pm(\bm{r})$ correspond to $\sigma_\pm$ components. There are three notable positions with high symmetry: $\bm{J}(\bm{r})$ has pure $\sigma_+$ component at $R_h^h$ sites, pure $\sigma_-$ component at $R_h^X$ sites, and vanishes at $R_h^M$ sites. Around the $R_h^h$ site ($\bm{r} = \bm{0}$), $J_+(\bm{r})$ is nearly a constant while $J_-(\bm{r})$ has a vortex structure: $J_-(\bm{r}) \propto r_x - i\, r_y$. Figures describing the spatial variation of $J_\pm(\bm{r})$ are provided in the supplementary information.

More than one *IX* resonance can be accommodated in a deep moiré potential as illustrated in Fig. 4a. Based on Eqn. (1), we calculate the theoretical optical conductivity of *K* valley *IX*s, as shown in Fig. 4b. Four resonances with *alternating* optical selection rules appear within the energy range of the potential. The calculated *IX* resonance lineshapes are homogeneously broadened and are significantly narrower than the inhomogeneous linewidths observed in our experiments as expected. Furthermore, the calculated *IX* peak height is proportional to the absorption coefficient while the measured PL intensity is determined by various relaxation channels. Thus, a mismatch in *IX* peak height between calculations and experiments is not surprising. The corresponding exciton COM wave function can be understood as Bloch wave states composed of Wannier functions confined to the potential minimum position ($R_h^h$ sites). In Figs. 4c-4e, we illustrate the COM wave functions of the first three resonances. Resonance (1) has *s*-wave symmetry centered at the $R_h^h$ site, and its optical matrix element has only $\sigma_+$ components. The wavefunction of resonance (2) is also centered at the $R_h^h$ site, but it has a chiral *p*-wave form with additional angular momentum. Due to this difference between the wave functions of resonances (1) and (2), their optical selection rules have opposite helicity. The behavior of resonances (3) and (4), which have *d*-wave and *f*-wave forms, can be understood in a similar way.

The assignment of the observed *IX* peaks as ground and excited states of excitons localized near the moiré potential minimum is consistent with the thermal behavior and recombination dynamics presented in Fig. 5. We show the steady-state PL at elevated temperatures between 25 K and 70 K in Fig. 5a. With increasing temperature, the rate at which the intensity of the two highest-energy peaks decreases is significantly faster than for the lower-energy peaks. Because excitons in the excited states are less-confined within the moiré pattern, they are more susceptible to phonon-induced activation out of the potential[39]. Excitons in the excited states can also relax to the lower energy states, which can enhance the recombination rate from these transitions. Indeed, we observe a faster decay of the excited states, shown by the time-resolved PL dynamics in Fig. 5b. The dynamics of the highest energy peak are fit by a single exponential with a 0.9 ns time constant. As the emission energy decreases, the dynamics become slower and biexponential, approaching decay times of 2 ns and 10 ns for the lowest energy state. The slight increase in the fast and slow decay times with decreasing energy, shown in the inset to Fig. 5b, is often observed in other systems with spatially localized excitons, such as in self-assembled InAs/GaAs quantum dots[40].

Other recent experiments have also reported multiple *IX* resonances. However, these experiments were performed on samples either with different stacking conditions or with significantly broader *IX* inhomogeneous linewidths that mask any effects of the moiré potential. Previous experiments have systematically investigated the large-twist angle dependence of PL emission from MoSe$_2$/WSe$_2$[41] and MoS$_2$/WSe$_2$[42] hBLs. In the case of MoSe$_2$/WSe$_2$, the emission intensity from the *IX* resonance is dramatically reduced as the twist angle deviates from the high-symmetry orientations of 0º or 60º because the lowest energy transitions become indirect in momentum space (*e.g.*, recombination along *Q-K* or *Γ-K*)[43]. In hBLs with *H*-type stacking (near 60° twist angle), two *IX* resonances separated by 25-50 meV have been assigned to such momentum-indirect transitions as reported by other groups[44,45] and also observed from an additional sample shown in the supplementary material. The assignment of momentum-indirect transitions is consistent with our calculations, which predict that the moiré potential in *H*-type hBLs is on the order of 10-20 meV. This potential depth is too small to lead to the two *IX* peaks with the observed energy splitting. For *R*-type hBLs, the direct *K-K* transition has a significantly larger dipole moment than the momentum-indirect transitions and thus is the dominant contribution to PL. The 1-2 ns lifetimes measured for our *R*-type sample are also consistent with the direct *K-K* transition, whereas momentum-indirect transitions in hBLs with larger twist angle exhibit recombination lifetimes of ~100 ns[32,46]. For momentum-indirect states, thermalization between the two transitions at elevated temperatures leads to weaker emission from the lower-energy resonance[45], in contrast to the thermal behavior of moiré excitons shown in Fig. 5a.

A recent theoretical study has also proposed *IX* resonances arising from $S_z = 1$ transitions, which are optically dark in monolayers but become bright in hBLs[47]. In an *R*-stacked heterostructure, $S_z = 1$ exciton recombination is predicted to emit left- (right-) circularly polarized light at the $R_h^X$ ($R_h^M$) atomic configurations. Since the $S_z = 1$ exciton at the *K* point consists of a spin-down conduction band electron and spin-up valence band electron (see Fig. 1b), it emits at an energy higher than that of the $S_z = 0$ states by the conduction band spin splitting of ~30 meV[48]. With increasing temperature, thermalization of excitons might lead to enhanced emission from $S_z = 1$ states, which is inconsistent with the temperature dependence of the excited states shown in Fig. 5a. The $S_z = 1$ states are expected to have longer recombination lifetimes than the $S_z = 0$ states due to a weaker transition dipole moment (see supplementary material), which is contrary to the trends in the measured lifetimes in Fig. 5b. Although we cannot completely rule out $S_z = 1$ states as a possible explanation for some of the observed resonances, such an explanation is less likely for the higher-energy states reported here, which are less stable states at higher temperature and exhibit a shorter lifetime.

From the future perspective of controlling moiré excitons, a natural choice would be to tune the moiré period through the twist angle. Indeed, in another hBL with small twist angle near 2°, we observe multiple *IX* resonances spaced by ~30 meV—larger than the spacing for the 1° sample as expected (see supplementary material). While systematic studies of the twist-angle dependence of *IX* in TMD hBLs have been performed for large (~5°) steps[41,42], inhomogeneous broadening has previously masked any effects of the moiré potential from being observed. Precise control of both the twist angle and the moiré potential depth is necessary to tune the energy spacing between the *IX* resonances in a predictable manner. An applied electric field or magnetic field may also allow one to control the *IX* properties. Although an electric field has been predicted to reverse the optical selection rules of moiré excitons[24], a rigid Stark shift is typically observed experimentally[44] without non-trivial changes in emission polarizations. Similarly, a magnetic field can be used to control spin states; however, the small Zeeman shift suggests that an exceptionally large magnetic field is required to introduce a shift comparable to the inhomogeneous linewidth of ~10 meV in available hBL samples[32,44]. With improved sample quality and reduced inhomogeneous broadening, we expect that the control of *IX* properties via external fields would be feasible.

In summary, we observed multiple interlayer exciton resonances in an hBN-encapsulated $MoSe_2$/$WSe_2$ heterobilayer. The key spectroscopic features observed in our experiments—four *IX* resonances nearly equally spaced in energy, alternating circularly polarized PL, systematic changes in the lifetime with energy, and the temperature dependence—cannot be naturally explained without invoking the moiré potential expected to exist in a stacked hBL. Because of the ~1 μm laser excitation spot size, our experiments are spatially averaged over many moiré supercells. Multiple moiré domains may exist within the excitation spot with slightly different twist angles, leading to inhomogeneous broadening of the optical resonances. The depth of the moiré potential depends on interlayer coupling, which is adjustable by choosing different materials, applying an electric field, or inserting a spacer layer. In the case of a deep moiré potential, this artificial lattice may serve as a uniform quantum dot array with exciting applications in quantum information technology for generating single photons. In self-assembled quantum dots extensively investigated for single-photon generation, the Coulomb force that binds the electron-hole pairs is treated as a perturbation to the confinement energy. The opposite situation is likely true in vdW heterostructures, where the quantum dot potentials from the moiré pattern can localize the exciton COM wavefunction. In the case of a small moiré period, exciton bands with novel topological properties may be engineered with an external electric and magnetic field[23,24]. Methods for loading a controlled number of excitons within each supercell may further allow interaction between excitons to be explored in a crystal with a period adjustable with the twist angle.

## Author Information

**Corresponding Author:** e-mail (X.Li): elaineli@physics.utexas.edu

**Notes:** The authors declare no competing financial interest.

**Acknowledgements:** K.T., J.C., A.H.M., and X.Li are partially supported by NSF MRSEC program DMR-1720595. L.S., A.Z., X.Li, C.S. are partially supported by NSF EFMA-1542747 and NSF DMR-1306878. X.Li also gratefully acknowledges the support from the Welch Foundation via grant F-1662. F.W. is supported by the Department of Energy, Office of Science, Materials Science and Engineering Division. X.Lu and L.Y. are supported by the Air Force Office of Scientific Research (AFOSR) FA9550-17-1-0304 and NSF DMR-1455346. K.W. and T.T. acknowledge support from the Elemental Strategy Initiative conducted by the MEXT, Japan and JSPS KAKENHI Grant No. JP15K21722. D.A.S acknowledges support from the NSF Graduate Research Fellowship Program. N.L. acknowledges support from NSF CMMI-1351875. K.K. and E.T. acknowledge support from Samsung Corp. The Texas Nanofabrication Facility where some of the work was done is a member of the National Nanotechnology Coordinated Infrastructure, which is supported by the National Science Foundation (Grant ECCS-1542159).

**Author Contributions:** K.T. and G.M. contributed equally to this work. K.T. performed the steady-state optical measurements. G.M. performed the time-resolved optical measurements. F.W. and X.Lu performed the calculations. K.T., J.C., J.E, A.Z., M.C., T.T. and K.W. prepared the samples. A.S., K.K., A.R., T.A. and D.A.S. assisted the experiments. K.T., G.M, F.W., and X.Li wrote the manuscript. E.T., S.B., N.L., K.L.S., L.Y., A.H.M., and X.Li supervised the project. All authors discussed the results and commented on the manuscript at all stages. This work is an official contribution of NIST, which is not subject to copyright in the United States.

## Methods

*Sample preparation:* The MoSe$_2$ and WSe$_2$ monolayers and hBN (hexagonal boron nitride) capping layers are prepared by mechanical exfoliation onto a polydimethylsiloxane (PDMS) sheet. The hBL is constructed layer-by-layer using the viscoelastic stamping method[29,30]. The bottom hBN is first transferred onto a sapphire substrate, followed by the WSe$_2$ monolayer. The top MoSe$_2$ monolayer is then stacked above the WSe$_2$ monolayer such that the crystal axes are rotationally aligned with a precision within 1º. Polarization- and phase-resolved second harmonic generation measurements are used to determine the crystal axes of the two monolayers[49-51]. Finally, a thin hBN layer (~15 nm thick) is deposit on top of the hBL. A microscopic picture of one sample is shown in Fig. 2a without the top hBN layer for clarity. The sample is annealed under ultrahigh vacuum (~$10^{-5}$ mbar) at 200 ºC for 8 hours.

*Optical spectroscopy:* For steady state PL measurements, the sample is held at 15 K and is excited using a continuous wave 660 nm laser focused to a spot size of 1.5 µm using a long working distance microscope objective with NA = 0.42. The transmitted PL signal is dispersed through a spectrometer and detected by a CCD. Time-resolved PL measurements were performed at a sample temperature of 5 K. A Ti:sapphire-pumped optical parametric oscillator emitting 100 fs, 680 nm pulses at a 76 MHz repetition rate was used as the excitation source. The PL was collected in the reflection geometry, spectrally filtered with 1 nm bandwidth using a monochromator, and time-resolved using a silicon single-photon avalanche detector. The time-resolved PL data shown in the main text were acquired using 100 µW average power. No change in the dynamics were observed for powers down to 1 µW.

**Data availability:** The data that support the findings of this study are available from the corresponding authors on reasonable request.

**Fig. 1. Moiré superlattice modulates the electronic and optical properties. (a)** Different local atomic alignments occur in an MoSe$_2$/WSe$_2$ vertical heterostructure with small twist angle. The three highlighted regions correspond to local atomic configurations with three-fold rotational symmetry. **(b)** In the $K$ valley, $S_z = 0$ interlayer exciton transitions occur between spin-up conduction-band electrons in the MoSe$_2$ layer and spin-up valence-band electrons in the WSe$_2$ layer. $K$-valley excitons obey different optical selection rules depending on the atomic configuration $R_h^\mu$ within the moiré pattern. $R_h^\mu$ refers to $R$-type stacking with the $\mu$ site of the MoSe$_2$ layer aligning with the hexagon center ($h$) of the WSe$_2$ layer. Exciton emission at the $R_h^h$ ($R_h^X$) is left-circularly (right-circularly) polarized. Emission from site $R_h^M$ is dipole-forbidden for normal incidence. **(c)** Left: The moiré potential of the interlayer exciton transition showing a local minimum at site $R_h^h$. Right: Spatial map of the optical selection rules for $K$-valley excitons. The high-symmetry points are circularly polarized, and regions between are elliptically polarized.

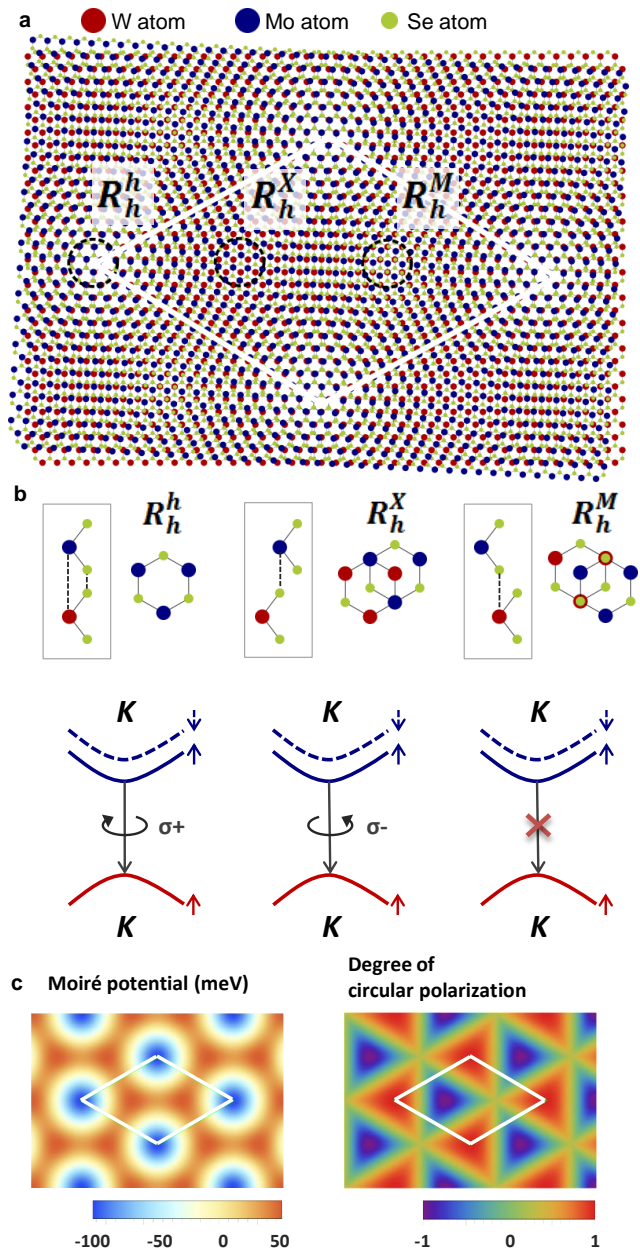

**Fig. 2. Photoluminescence from MoSe$_2$/WSe$_2$ heterobilayer.** (a) Optical image of an hBN-encapsulated MoSe$_2$/WSe$_2$ stacked heterostructure. The hBL region is indicated inside the black dotted line. (b) Comparison of the photoluminescence spectrum from an uncapped heterostructure (dashed curve) and an hBN-encapsulated heterostructure (solid curve). Neutral ($X^0$) and charged ($X^-$) exciton emission is observed from the MoSe$_2$ and WSe$_2$ monolayers. The interlayer exciton ($IX$) emission is observed ~200 meV below the intralayer resonances. (c) Illustrative band diagram showing the type-II alignment and the $IX$ transition.

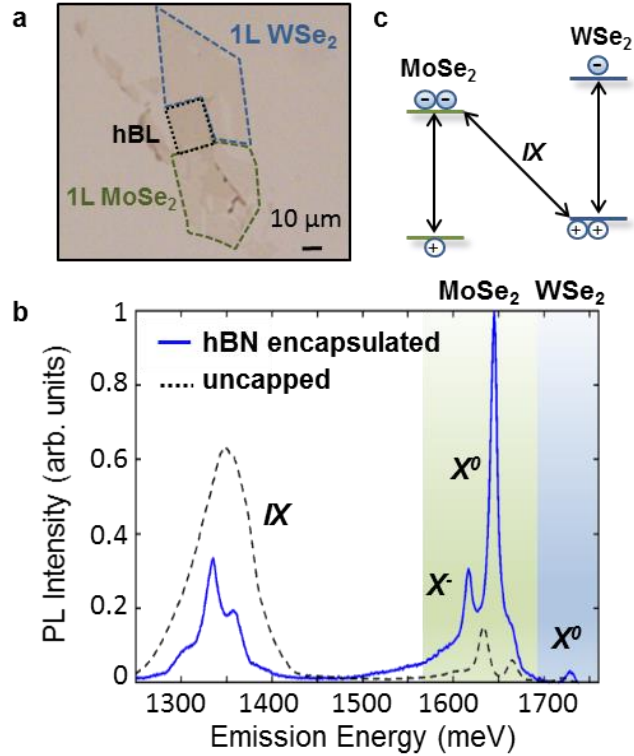

**Fig. 3. Circularly polarized emission. (a)** Spatial dependence of the PL spectra shown for three different positions across the hBL sample. The spectrum is fit with four Gaussian functions at each position. **(b)** The center energy of each peak obtained from the fits at different sample positions. **(c)** Circularly polarized PL spectrum for σ+ excitation. **(d)** The degree of circular polarization versus emission wavelength obtained from the spectra in **(c)**.

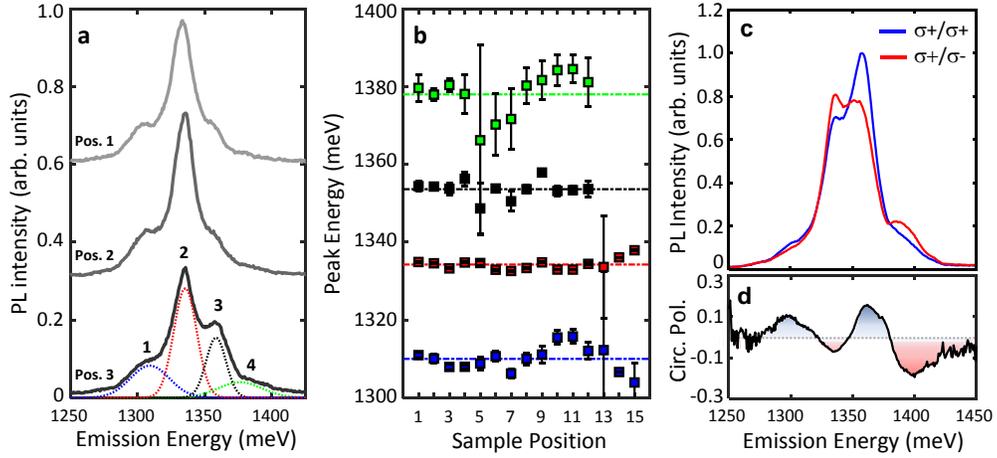

**Fig. 4. A quantum dot array defined by the moiré superlattice.** (a) Illustration of the spatial variation of the moiré potential and the multiple confined *IX* resonances. (b) Optical conductivity of *IX*s in the *K* valley in response to $\sigma_+$ (blue line) and $\sigma_-$ (red line) polarized light. **(c)-(e)** Real-space map of the center-of-mass wave functions for peaks (1), (2), and (3), respectively.

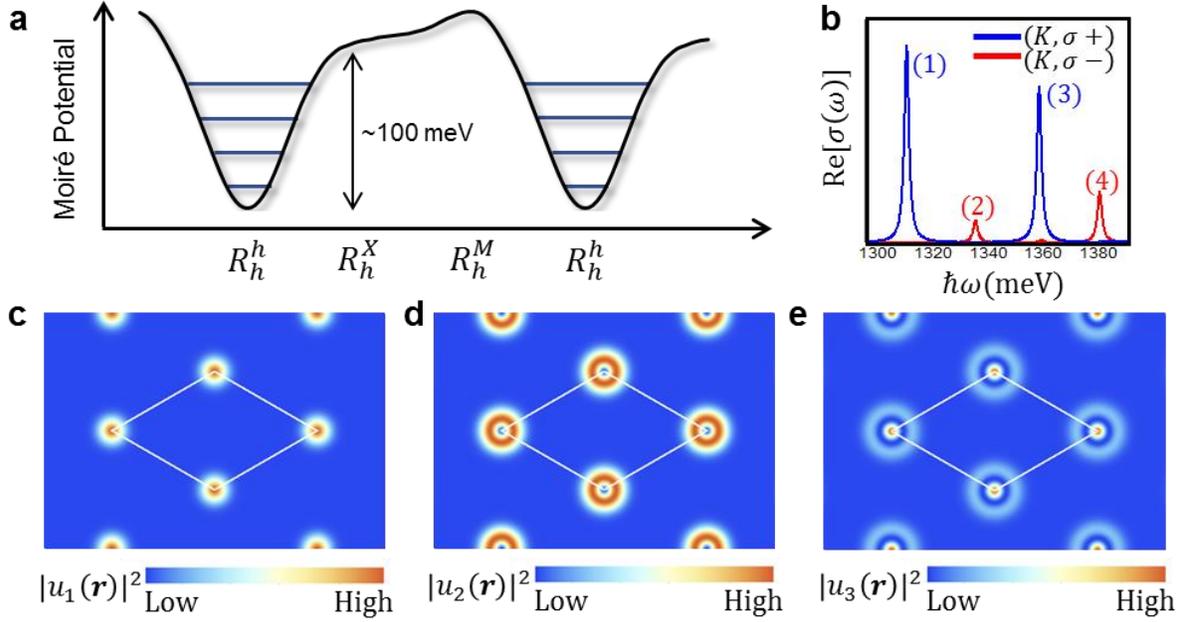

**Fig. 5. Thermal decay and recombination dynamics. (a)** Temperature dependence of the PL between 25 K and 70 K. **(b)** Time-resolved PL dynamics (points) at energies near the four *IX* transitions labeled in the inset. The solid lines are biexponential fits to the data. The inset shows the emission energy dependence of the fast and slow decay times.

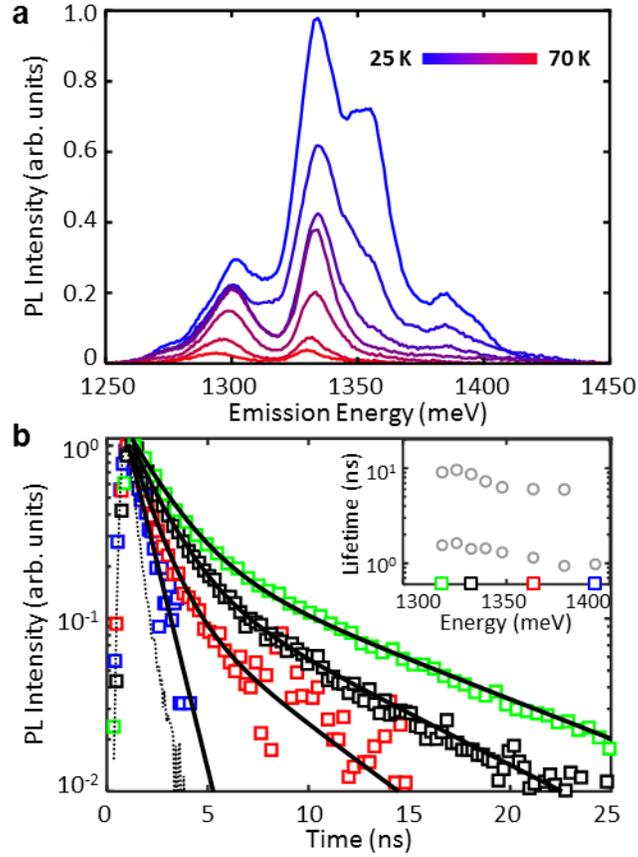